\documentclass{ws-procs10x7}
\usepackage{balance}

\newcolumntype{d}[1]{D{.}{.}{#1}}

\def\Journal#1#2#3#4{{\it #1} {\bf #2}, #3 (#4)}

\makeindex
\begin{document}

\title{NEMO-3 DOUBLE BETA DECAY EXPERIMENT: LASTEST RESULTS}

\author{A. S. BARABASH$^*$ (on behalf of the NEMO Collaboration)}

\address{Institute of Theoretical and Experimental Physics, Moscow, 117218, 
Russia\\$^*$E-mail: barabash@itep.ru}


\twocolumn[\maketitle\abstract{Latest results on $0\nu\beta\beta$, 
$0\nu\chi^0\beta\beta$ and $2\nu\beta\beta$ 
   decays 
of different isotopes from NEMO-3 double beta decay experiment are presented. In particular, 
new limits at 90\% C.L. on neurtinoless
double beta decay of $^{100}Mo$ and $^{82}Se$ have been obtained, $T_{1/2} > 5.8\times 10^{23}$ y and  
$T_{1/2} > 2.1\times 10^{23}$ y, respectively.}
]

\section{Introduction}

The experiments with solar, atmospheric, reactor and accelerator neutrinos have provided compelling 
evidences for the existence of neutrino oscillations driven by nonzero neutrino masses 
and neutrino mixing (see recent reviews \cite{VAL06,BIL06,MOH06} and reference therein).
However, the experiments
studying neutrino oscillations are not sensitive to the nature of
the neutrino mass
(Dirac or Majorana?) and provide no information on the absolute
scale of the neutrino
masses, since such experiments are sensitive only to the $\Delta m^{2}$. 
The detection
and study of $0\nu\beta\beta$ decay may clarify the following problems
of neutrino physics
(see discussions in Ref. \cite{PAS03,PAS06}): (i) neutrino
nature; is the neutrino a
Dirac or a Majorana particle?, (ii) absolute neutrino mass scale
(a measurement or a limit
on m$_1$), (iii) the type of neutrino mass hierarchy (normal,
inverted, or  quasidegenerate),
 (iv) CP violation in the lepton sector (measurement of the
Majorana CP-violating phases).

In connection with the $0\nu\beta\beta$ decay, the detection of double beta decay 
with the emission of two neutrinos ($2\nu\beta\beta$), which is allowed process 
of second order in the Standard Model, enables the experimental determination 
of nuclear matrix elements involved in the double beta decay processes. Accumulation 
of experimental information for the $2\nu\beta\beta$ processes (transitions to the 
ground and excited states) promotes a better understanding of the nuclear part of 
double beta decay, and allows one to check theoretical schemes of nuclear matrix 
element (NME) calculations for the two neutrino mode as well as for the neutrinoless one.

The main objective of the NEMO-3 experiment is to seek the $0\nu\beta\beta$ decay 
of various isotopes ($^{100}$Mo, $^{82}$Se, etc.) with a sensitivity to half-life 
up to $\sim n\times 10^{24}$ y.
In addition, one of the tasks consists in studying, at a high level of precision, the
$2\nu\beta\beta$ decay of a broad range of nuclei ($^{100}$Mo, $^{82}$Se, $^{116}$Cd, $^{150}$Nd,
$^{130}$Te, $^{96}$Zr and $^{48}$Ca) and currently exploring all features of double beta decay
processes.

The NEMO-3 experiment is a tracking experiment where, in contrast to experiments with $^{76}$Ge 
\cite{KLA01,IGEX02},
one detect not only the total energy deposition but also the remaining parameters of the process, 
including the energy of individual electrons, their divergence angle, and the coordinate of an 
event in the source plane. Since June 2002, the NEMO-3 detector has operated at the 
Frejus Underground Laboratory (France) located at a depth of 4800 m w.e. Since February 2003,
the detector has been routinely taking data.

\section{The NEMO-3 detector}

The detector has a cylindrical
structure and consists of 20 identical sectors
A thin (about 30-60 mg/cm$^{2}$) source containing double beta-decaying
nuclei and having a total area of 20 m$^{2}$ and a weight
of up
to 10 kg was placed in the detector. The energy of the electrons is measured by plastic
scintillators (1940 individual counters), while the tracks
are reconstructed on the basis of information obtained in the
planes of Geiger cells (6180 cells) surrounding the source
on both sides. The tracking volume of the detector is filled with
a mixture consisting of $\sim$ 95\% He, 4\% alcohol, 1\% Ar
and 0.1\% water at slightly above atmospheric
pressure. In addition, a magnetic field of strength of about 25 G
that is parallel to the detector axis is created by a solenoid
surrounding the detector. The magnetic field is used to identify
electron-positron pairs to suppress this
source of background.

The main characteristics of the detector are the following: the
energy resolution of the scintillation counters lies in
the interval 14-17\% FWHM for electrons of energy 1 MeV; the time
resolution is 250 ps for an electron energy of
1 MeV; and the accuracy in reconstructing of the vertex of 2e$^{-
}$ events is about 1 cm.
The detector is surrounded by a passive shield consisting of 20 cm
of steel and 30 cm of borated water. The level of
radioactive impurities in structural materials of the detector and
of the passive shield was tested in measurements
with low-background HPGe detectors.

Measurements with the NEMO-3 detector revealed that tracking
information, combined with time and energy measurements,
makes it possible to suppress the background efficiently. That
NEMO-3 can be used to investigate almost all isotopes
of interest is a distinctive feature of this facility. At the
present time, such investigations are being performed
for seven isotopes; these are $^{100}$Mo (6.9 kg), $^{82}$Se (0.93 kg), 
$^{116}$Cd (0.4 kg),
$^{150}$Nd (36.6 g), $^{96}$Zr (9.4 g), $^{130}$Te (0.45 kg), and $^{48}$Ca 
(7 g). In addition, foils from copper and natural (not
enriched) tellurium are placed in the detector for performing
background measurements. A detailed description of the detector
and its characteristics is presented in Ref. \cite {ARN05}.

After starting the experiment, a small amount of radon was detected inside the
tracking chamber. It was demonstrated that the major contribution was coming from laboratory air
through small leaks. To solve the problem a tight tent surrounding the
detector was built. The tent is fed with radon free air, coming from a
radon trapping facility. The core of this facility is a charcoal tank operating
at -50 $^o$C. Radon is trapped inside the charcoal and decays while
diffusing through it. So no regeneration is needed.

A reduction factor of 100 was achieved for the radon in the air inside the tent. 
The radon level in the
tracking chamber dropped from 20-30 mBq/m$^3$ to 4-5 mBq/m$^{3}$. It is now the case
that the remaining radon in the detector is dominated by the degassing of the 
construction components.

Thus there are two data sets. Phase I data corresponding to a higher radon background,
and Phase II (since December 2004), with low background conditions.

\section{Experimental results}

\subsection{$^{100}$Mo results}

\begin{figure}
\includegraphics[width=7cm]{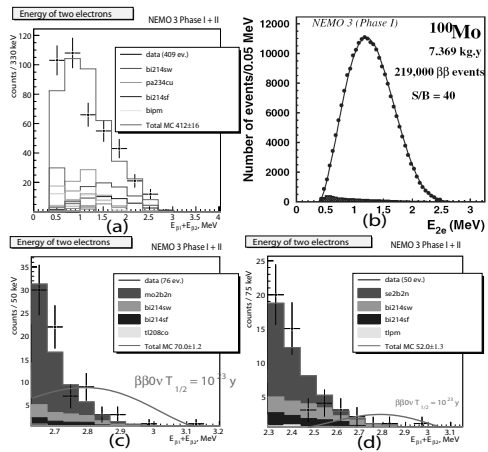}
\caption{(a) Background $\beta\beta$
spectrum in Cu foil (Phase I + II); (b) $^{100}\rm Mo$ Phase I
$2\nu\beta\beta$ spectrum; (c) $^{100}\rm Mo$ (Phase I + Phase II)
$2\nu\beta\beta$ spectrum at $Q_{\beta\beta}$; (d) $^{82}\rm Se$ (Phase I + Phase II)
$2\nu\beta\beta$ spectrum at $Q_{\beta\beta}$.}
\label{fig1}
\end{figure}

A measurement of $2\nu\beta\beta$ decay was done with Phase I data, see
Table \ref{tb_2b2n} and Fig. \ref{fig1}(b)
\cite{ARN05a}. Given the largest statistical data set  in the world
(219000 events) and a signal-to-background ratio
(S/B) of 40, the single state dominance (SSD) mechanism \cite{SIM01} of the decay was
confirmed.
Also $^{100}\rm Mo$ $2\nu\beta\beta$ decay to the excited
$0^+ $ state (1130.29 keV) of $^{100}\rm Ru$ was measured with a half life of
$T_{1/2}=[5.7 ^{+ 1.3} _{-0.9}(\rm stat) \pm 0.8 (\rm syst)] \times 10^{20}$ years.
All these results are important input for nuclear physics theory. They can be used to
validate nuclear model for further $0\nu\beta\beta$ nuclear matrix elements
calculations.

In the $0\nu\beta\beta$ decay search, no signal was found ( Fig.
\ref{fig1}(c)).
The data for Phase I and Phase II were combined. A preliminary
counting analysis shows 14 events in the window of interest [2.78-3.20]
MeV, the expected background is 13.4 events, and $\beta\beta_{0\nu}$
efficiency is 8.4\%. The effective time analyzed is 13 kg$\cdot$y yielding a
lower limit on the half-life of $T_{1/2} > 5.8 \times 10^{23}$ y (90\% C.L.),
corresponding to the effective neutrino mass $\langle m_{\nu} \rangle < 0.6-0.9$ eV
using NME from Ref. \cite{SIM99,STO01,CIV03} or $\langle m_{\nu} \rangle < 2.0-2.7$ eV
using recent value of NME from Ref. \cite{ROD06}.

In the hypothesis of a right-handed weak current, the limit is 
$T_{1/2} > 3.2 \times 10^{23}$ y (90\% C.L.), corresponding to an upper limit
of the coupling constant of $\lambda < 1.8 \times 10^{-6}$ using NME from 
Ref. \cite{SUH02}.

\subsection{$^{82}$Se results}

2570 $2\nu\beta\beta$ events were registered during Phase I of the
experiment, with a S/B ratio equal to  about 4, and a half-life given in Table \ref{tb_2b2n}.

After the preliminary analysis of 1.76 kg$\cdot$y of Phase I and Phase II data,
see Fig. \ref{fig1}(d), 7
events were found in the window [2.62-3.20] MeV, with the expected background 6.4
events, $0\nu\beta\beta$ efficiency of 14.4\% yielding a lower limit on the half-life of
$T_{1/2}>2.1\times 10^{23}$ y (90\% C.L.), which corresponds to an upper
mass limit of $\langle m_{\nu} \rangle < 1.2-2.5$ eV using NME from Ref. \cite{SIM99,STO01,CIV03} 
or $\langle m_{\nu} \rangle < 2.3-3.2$ eV
using recent value of NME from Ref. \cite{ROD06}.

In the hypothesis of a right-handed weak current, the limit is 
$T_{1/2} > 1.2 \times 10^{23}$ y (90\% C.L.), corresponding to an upper limit
of the coupling constant of $\lambda < 2.8 \times 10^{-6}$ using NME from 
Ref. \cite{AUN98}.

\subsection{$2\nu\beta\beta$ decay of other nuclei}

Phase I data for four other isotopes ($^{116}$Cd, $^{150}$Nd,
$^{96}$Zr and $^{48}$Ca) were analyzed and their half-lives
measured, see Table \ref{tb_2b2n} with preliminary
results. This should be a very important guide for
nuclear theory. Since all isotopes are measured with the same device and at the same time, the
half-life ratio has a very small systematic uncertainty, while statistical errors will
reach a few per cent.

\begin{table*}

\tbl{Main results on $2\nu\beta\beta$ decays.\label{tb_2b2n}}
{\begin{tabular}{cccc}
\toprule
Nuclei & Measurement & Number of & $T_{1/2}$, y\\
  & time, days & $2\nu$ events &  \\
\colrule
$^{100}\rm Mo$ & 389 & 219000 & $[7.11 \pm 0.02 (\rm stat) \pm 0.54(\rm syst)] \times 10^{18}$ \\
$^{100}\rm Mo$-$^{100}\rm Ru(0^{+}_{1})$ & 334.3 & 38 & 
$[5.7 ^{+ 1.3} _{-0.9}(\rm stat) \pm 0.8 (\rm syst)] \times 10^{20}$ \\ 
$^{82}\rm Se$ & 389 & 2750 & $[9.6 \pm 0.3 (\rm stat) \pm 1.0(\rm syst)] \times 10^{19}$ \\
$^{116}\rm Cd$ & 168.4 & 1371 & 
$[2.8 \pm 0.1 (\rm stat) \pm 0.3(\rm syst)] \times 10^{19}$ \\
$^{150}\rm Nd$ & 168.4 & 449 & 
$[9.7 \pm 0.7 (\rm stat) \pm 1.0(\rm syst)] \times 10^{18}$ \\
$^{96}\rm Zr$ & 168.4 & 72 & 
$[2.0 \pm 0.3 (\rm stat) \pm 0.2(\rm syst)] \times 10^{19}$ \\
$^{48}\rm Ca$ & 446.7 & 51 & 
$[3.9 \pm 0.7 (\rm stat) \pm 0.6(\rm syst)] \times 10^{19}$ \\
\botrule
\end{tabular}}
\end{table*}

\subsection{Neutinoless double beta decay with Majoron emission}

8023 h of Phase I data were analyzed. The half-life limits for $^{100}$Mo 
and $^{82}$Se for the different modes with different value of spectral index n 
are presented in Table \ref{tb_maj} (see details in Ref. \cite{ARN06}). In particular,
limits on "ordinary" Majoron with n = 1 are $T_{1/2} > 2.7 \times 10^{22}$ y (90\% C.L.)
for $^{100}$Mo and $T_{1/2} > 1.5 \times 10^{22}$ y (90\% C.L.) for $^{82}$Se.
Using NME from Ref. \cite{SIM99,STO01,CIV03} one can obtain limits on coupling 
constant of Majoron to neutrino $\langle g_{ee} \rangle < (0.4-0.7) \times 10^{-4}$ 
and $< (0.7-1.4) \times 10^{-4}$ respectively. If one uses NME from Ref. \cite{ROD06}
then limits are $< (1.3-1.8) \times 10^{-4}$ for $^{100}$Mo and 
$< (1.3-1.8) \times 10^{-4}$ for $^{82}$Se.

\begin{table*}
\tbl{Limits (in years) on $0\nu\beta\beta$ decay with Majoron emission for $^{100}$Mo 
and $^{82}$Se. n - spectral index. \label{tb_maj}}
{\begin{tabular}{@{}ccccc@{}}
\toprule\\[-6pt]
Nuclei & n =1 & n = 2 & n = 3 & n = 7 \\
\colrule  
$^{100}\rm Mo$ &  $> 2.7 \times 10^{22}$ & $> 1.7 \times 10^{22}$ 
& $> 1.0 \times 10^{22}$ & $> 7.0 \times 10^{19}$ \\ 
$^{82}\rm Se$ &  $> 1.5 \times 10^{22}$ & $> 6 \times 10^{21}$ 
& $> 3.1 \times 10^{21}$ & $> 5.0 \times 10^{20}$ \\ 

\botrule
\end{tabular}}
\label{tb_maj}
\end{table*}

\section{Conclusion}

The NEMO-3 detector has been running reliably since February 2003 and 
since December 2004 under "low Rn" background conditions. The $2\nu\beta\beta$
decay has been measured for $^{48}$Ca, $^{82}$Se, $^{96}$Zr, $^{100}$Mo,
$^{116}$Cd, and $^{150}$Nd. In addition $2\nu\beta\beta$ decay of $^{100}$Mo to the $0^+$
(1130.29 keV) excited state of $^{100}$Ru has been investigated. No evidence for $0\nu\beta\beta$ is found.

At present, NEMO Collaboration continue data analysis and new results will be obtained soon. 
The NEMO-3 detector is continuing to collect data and for 5 y of 
measurements the sensitivity of experiment for $0\nu\beta\beta$ decay will be increased up to
$\sim 2\times 10^{24}$ y for $^{100}$Mo and $\sim 8\times 10^{23}$ y for $^{82}$Se.

\section*{Acknowledgments}

A portion of this work was supported by grant from INTAS (03051-3431).

\end{document}